# Application of Machine Learning for Online Dynamic Security Assessment in Presence of System Variability and Additive Instrumentation Errors

Anubhav Nath, *Member, IEEE*, Reetam Sen Biswas, *Student Member, IEEE,* and Anamitra Pal, *Member, IEEE*
School of Electrical, Computer, and Energy Engineering
Arizona State University, Tempe, Arizona-85287, USA
Email: anath7@asu.edu, rsenbisw@asu.edu, anamitra.pal@asu.edu

*Abstract*—Large-scale blackouts that have occurred in the past few decades have necessitated the need to do extensive research in the field of grid security assessment. With the aid of synchrophasor technology, which uses phasor measurement unit (PMU) data, dynamic security assessment (DSA) can be performed online. However, existing applications of DSA are challenged by variability in system conditions and unaccounted for measurement errors. To overcome these challenges, this research develops a DSA scheme to provide security prediction in real-time for load profiles of different seasons in presence of realistic errors in the PMU measurements. The major contributions of this paper are: (1) develop a DSA scheme based on PMU data, (2) consider seasonal load profiles, (3) account for varying penetrations of renewable generation, and (4) compare the accuracy of different machine learning (ML) algorithms for DSA with and without erroneous measurements. The performance of this approach is tested on the IEEE-118 bus system. Comparative analysis of the accuracies of the ML algorithms under different operating scenarios highlights the importance of considering realistic errors and variability in system conditions while creating a DSA scheme.

*Index Terms*—Dynamic Security Assessment (DSA), Machine Learning (ML), Phasor Measurement Unit (PMU), Renewable Generation.

## I. Introduction

The power systems of the modern world need an improved situational awareness scheme that enables their operators to have better visualizations of their systems. DSA was designed to satisfy this need. Ref. [1] defines DSA as follows: *"DSA refers to the analysis required to determine whether or not a power system can meet specified reliability and security criteria in both transient and steady-state time frames for all credible contingencies."* DSA mainly deals with transient stability and/or short-term voltage stability and does an assessment of the power system's ability to maintain synchronism when the system is experiencing sudden disturbances such as loss of loads, stalling of motors, or short circuit on transmission lines [2]. A post-fault DSA scheme kicks into action after a fault is detected in the system. However, if the fault is not detected in time, this scheme would fail, causing unforeseen consequences. A pre-fault DSA scheme is fault independent and can mitigate the risk of cascading blackouts by alerting the operator to possible *preventive control actions* that can be undertaken. Thus, DSA scheme with pre-fault capability is the need of the hour, and the focus of this paper.

In [3], Luo used a fuzzy classification method to determine dynamic security. Direct methods for transient stability assessment employing Lyapunov stability were utilized in [4]-[7]. A variety of PMU-based indices for DSA including those for short-term voltage stability were explored in [8]. Beside the above-mentioned approaches, probabilistic methods, and dynamic state estimation have also been used for performing DSA [9]-[12].

With the advent of PMUs, large amounts of data at higher resolution and fidelity, have become available. Data mining techniques employing advanced learning methods can make use of this data to draw hidden inferences. In [13], Sun et al. proposed the use of decision tree (DT) to perform DSA. DT based preventive and corrective applications for DSA was also proposed in [14]. Random forest (RF) was used in [15] to classify the security status of DSA. Adaboost classifiers were used in [16] to determine DSA using PMU data. In [17], an extreme learning machine (ELM) based DSA scheme was developed that took into consideration large penetration of wind energy. Support vector machine (SVM) was used in [18]-[19] to assess system security.

However, [13]-[19] have not considered realistic errors in PMU measurements as well as the load variations that can occur in a particular season of the year, both of which must be taken into consideration when performing DSA. Furthermore, due to the increasing amount of renewable generation in the grid, it is important to also incorporate varying levels of renewable penetration while performing DSA studies. These knowledge gaps identified in the literature have been addressed in this paper using different ML techniques.

The rest of the paper is structured as follows. In Section II, the proposed DSA scheme has been explained. In Section III, the different ML algorithms employed in the study have been summarized. In Section IV, a case study on the IEEE-118 bus system has been presented. In Section V, we discuss the simulation results and the effectiveness of the approach. The conclusion is provided in Section VI.

## II. ONLINE DYNAMIC SECURITY ASSESSMENT (DSA) SCHEME

Online DSA plays an important role in determining the security of the power system in real-time. It assists the power system operator in operational decision-making and initiating remedial control processes. The flowchart for the proposed ML based online DSA scheme is shown in Fig. 1. The proposed approach is executed in the following three stages.

### A. Stage 1: Offline ML technique building

In this stage, multiple operating conditions, $N_{oc}$, are generated on a season-based load profile. The different operating conditions are obtained through different combinations of load and generation for the given season. For each $N_{oc}$, time domain simulation (TDS) of $N_C$ contingencies are executed. Next, specified security criteria, dealing with transient stability and short-term voltage security are checked to determine the security classification for each case. In this study, the classification parameter is binary ("1" for a secure case and "0" for an insecure case). Finally, a database of $N_{oc} \times N_C$ cases is generated which contains a security classification along with a vector of predicted values. 70% of the database is used for training while the remaining 30% is used for testing. Additionally, 10% of the training data is used for validation. Different ML algorithms are trained on the created database for each season to create the trained model. '$N-1$' contingencies and multiple '$N-k$' contingencies are simulated to create the secure and insecure cases. The simulation length is 20 seconds, with the contingencies executed at the $5^{th}$ second. All contingencies are three-phase line to ground faults located at 10% of the distance of the line from the "from" bus. The largest contingency simulated was the simultaneous opening of 6 lines (i.e. $k \leq 6$).

### B. Stage 2: Select ML model

In this stage, the ML model that gives the best results is identified. For the four seasons (summer, fall, winter, and spring), the ML models-under study are fed into the online DSA scheme. The performance classifiers for each of the ML models are tested to determine which one gives best performance. The simulations are repeated multiple times and a 95% confidence interval is used to account for the deviation in the performance of the ML algorithms due to different system conditions.

### C. Stage 3: Perform online DSA

In real-time, the control centers obtain synchronized measurements from the PMUs to perform DSA for single or multiple contingencies. In this study, it has been assumed that the online measurements are obtained from PMUs *only*. A PMU's sampling rate is high (30 samples/second) and this research utilizes this high sampling rate to determine the security/insecurity of the operating conditions in real-time. A window of 30 samples is selected by the proposed DSA scheme to ascertain the security of the system. The operator will employ the optimal ML model identified in Stage 2 of the assessment scheme for the online implementation.

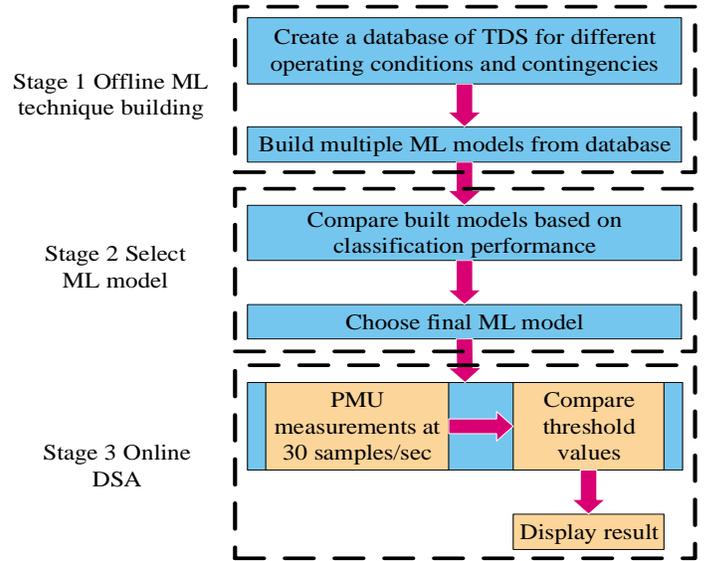

Fig. 1. The flowchart for the PMU and ML based online DSA scheme

## III. OVERVIEW OF MACHINE LEARNING ALGORITHMS

In this study, four different ML algorithms have been employed to classify the security or insecurity of the operating conditions of the power system, namely, decision tree (DT), random forest (RF), support vector machine (SVM), and multi-layer neural network (MLNN).

DT has been one of the most popular classification algorithms in power systems and has been extensively used for performing DSA [13]-[14], [20]-[21]. In this study, a classification and regression tree (CART) based DT has been trained offline with the help of a training database and a DT model has been developed by identifying correlations between the input and the output. RF is an ensemble learning technique of classification or regression that operates by constructing a multitude of decision trees during training phase and subsequently outputting the class that is the mode of the classes (for a classification problem) or the mean prediction (for a regression problem) of the individual trees [22]. RF has been used for performing DSA in [15]. SVMs are machine learning models based on statistical learning theory [23] and have been used for security assessment in [18]-[19], [24]. The MLNN used in this paper comprises of a feed-forward neural network called a multilayer perceptron (MLP), with a self-exponential linear unit (SELU) as the activation function [25]-[26]. SELU has self-normalizing properties because the activations that are close to zero mean and unit variance, when propagated through many network layers, converge to zero mean and unit variance even if noise is present in the data. This improves the robustness of the algorithm. To the best of the authors' knowledge, [27] is the only article that explores the use of neural networks for performing DSA using PMU data.

## IV. CASE STUDY: IEEE 118 BUS SYSTEM

To verify the performance of the proposed DSA technique, simulations were carried out on the IEEE-118 bus. The system consists of 118 buses, 54 generators, 177 transmission lines, and 9 transformers [28].

## A. Incorporation of Seasonal Load

In this study, an attempt has been made to segregate the year into 4 seasons, namely, spring, summer, fall, and winter to create four normative load profiles that can more accurately represent the load variations for different seasons. In [29], California independent system operator (CAISO) has uploaded the hourly load profile for its energy management system (EMS) for the years 2014-2017. Based on the hourly load for each of the season, an average has been taken for each of the 24 hours to find the net load curve representing a 24-hour period that would best represent each season. The process is repeated for 4 years' worth of data to account for any load change that might have happened over the years. The summer load profile for the aggregate over the four years has been shown in Fig. 2.

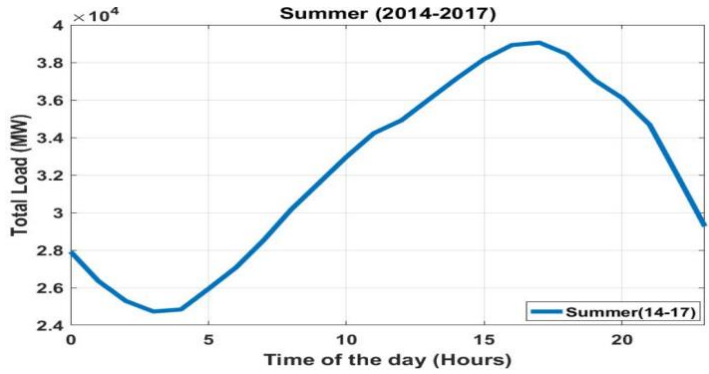

Fig. 2. Summer daily load in MW

## B. Simulation setup

Time domain simulation (TDS) is carried out on a computer with Intel Core i7-6700 CPU @ 3.4Ghz with 16GB RAM. The database not only contains the measurement data which are the predictor values, but also the target values, namely, secure (1) and insecure (0). The latter is assigned according to the following criteria:

a) Transient Stability: The system is termed transient stable for a given contingency if the transient stability index (TSI) [30] of the system is lower than 10% [31]. Mathematically, TSI can be expressed as,

$$TSI = \frac{360 - \delta_{max}}{360 + \delta_{max}} * 100\% \quad (1)$$

In (1), $\delta_{max}$ is the maximum angle separation of any two rotor angles in degrees. Note that the TSI calculated using (1) is based on the maximum power swing algorithm.

b) Short-term Voltage Stability: A system is said to suffer from short-term voltage insecurity if the duration of any bus voltage is outside the range of 0.8 p.u. and 1.1 p.u. for more than 0.5 seconds [31].

To create the database, both secure and insecure cases have been simulated using TSAT software [30] by following four steps: i) Generation of cases; ii) Measurement of voltage magnitude and voltage angles from optimally placed PMUs; iii) Building of the training database containing predictor values; iv) Implementation of the ML algorithms on the testing database. To test the ML models summarized in Section III, realistic measurements are created by introducing measurement errors in the training database of true voltage phase angles and true voltage phase magnitudes. In accordance with [32]-[33], additive error model is used which includes both PMU and instrumentation channel errors, as described below:

a) PMU errors in phase angles are assumed to follow a Gaussian distribution with zero mean and standard deviation of $0.104°$ while the errors in phase magnitude are assumed to have a Gaussian distribution having zero mean and standard deviation of 0.15%.

b) Instrumentation channel errors in phase angle are assumed to have a uniform distribution in the range of $\pm 3°$, $\pm 2°$ and $\pm 1°$ while the errors in phase magnitude are assumed to follow a uniform distribution having zero mean and standard deviation of 0.20% [34].

The resultant voltage phase angles and magnitudes after inclusion of additive PMU and instrumentational channel errors are given by:

$$\emptyset_{actual}^v = \emptyset_{true}^v + \emptyset_{error}^c + \emptyset_{error}^{PMU} \quad (2)$$
$$V_{actual}^m = V_{true}^m + V_{error}^c + V_{error}^{PMU} \quad (3)$$

where $\emptyset_{actual}^v$ and $V_{actual}^m$ is the resultant voltage phase angle and magnitude after incorporation of errors in the true voltage phase angle and magnitude measurements $\emptyset_{true}^v$ and $V_{true}^m$ The instrumentation channel error is $\emptyset_{error}^c$ and $V_{error}^c$ while the PMU error is $\emptyset_{error}^{PMU}$ and $V_{error}^{PMU}$

## C. Importance of considering seasonal load modeling

In this study, varying load profiles that occur in different seasons have been considered. Consider the following case-study where a $N-3$ contingency (3 three-phase line to ground faults) is initiated at 5 seconds and the simulation is run for 20 seconds. It can be observed from Fig. 3 that for a summer load profile, the rotor angle of generator 40 is swinging away from the system and the simulation lasts for only 8.6 seconds, implying that the system is becoming unstable due to violation of TSI. However, the system is transient stable for the winter load profile for the same contingency case (see Fig. 4). This case-study emphasizes the need for doing a season-based load modeling while performing DSA studies.

## D. Renewable generation modeling and its importance

In this study, varying levels of solar penetration have been considered while performing DSA. Consider the following case-study where a $N-2$ contingency (2 three-phase line to ground faults) is initiated at 5 seconds for two different system conditions of the IEEE-118 bus system for a summer season load profile. In the first scenario, no solar PV is present in the system, whereas in the second scenario, solar PV is installed in the system by replacing the conventional generation at bus number 54. Fig. 5 represents the bus voltage magnitude of bus number 56 for the system without solar PV while Fig. 6 represents the bus voltage magnitude of bus number 56 when the solar PV is present. We observe that following a contingency, the voltage at bus number 56 does not suffer any short-term voltage violation and rises to a stable value after the initial dip when solar generation is not present (Fig. 5). Conversely, when the same contingency happens in the system with solar PV added, there is a short-term voltage violation at bus number 56 (Fig. 6) and the simulation lasts only till 5.5 seconds. This case-study highlights the need for performing DSA studies while considering different percentages of renewable generation that a system may be subjected to.

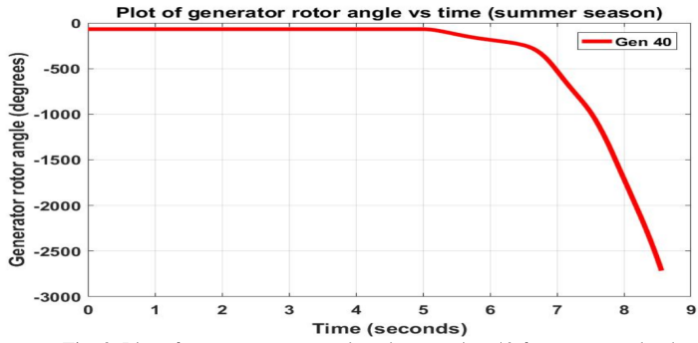
Fig. 3. Plot of generator rotor angle at bus number 40 for a summer load profile

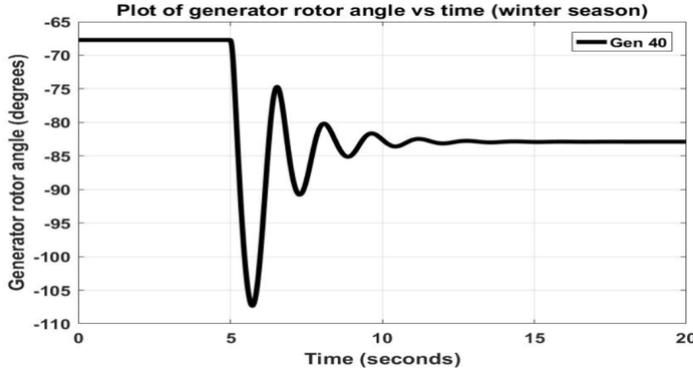
Fig. 4. Plot of generator rotor angle at bus number 40 for a winter load profile

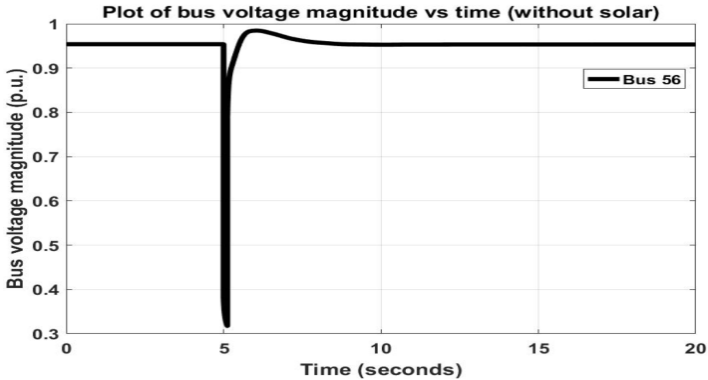
Fig. 5. Plot of bus voltage magnitude at bus number 56 (without solar PV) for a summer load profile

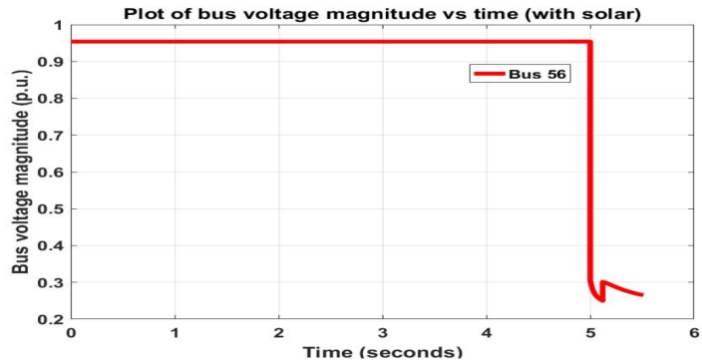
Fig. 6. Plot of bus voltage magnitude at bus number 56 (with solar PV) for a summer load profile

*E. PMU Placement*

The objective of the PMU placement problem is to guarantee observability of the system with minimum number of PMUs. There have been multiple PMU placement techniques such as the ones proposed in [35]-[38]. An integer programming formulation is used in this paper which is based on the methodology presented in the above publications to compute the PMU locations. The proposed logic for PMU placement can be mathematically expressed as,

$$min. \sum_{i=1}^{n} c_i . y_i \quad (3)$$
$$s.t\ f(Y) \geq \hat{a} \quad (4)$$

where $c_i$ is the cost of the placement of a PMU at bus $i$, $\hat{a}$ is the $n \times 1$ vector having all ones as its entries and $Y$ is the vector which is binary indicating placement of a PMU. The entries of the binary vector $Y$ have been defined as follows:

$$y_i = \begin{cases} 1, & if\ a\ PMU\ has\ been\ installed\ at\ bus\ 1 \\ 0, & otherwise \end{cases} \quad (5)$$

The binary incidence matrix $M$ is used to represent the system connection configuration having entries as follows:

$$M_{i,j} = \begin{cases} 1, & if\ i\ and\ j\ are\ connected\ or\ i = j \\ 0, & otherwise \end{cases} \quad (6)$$

To guarantee full observability of the system, each bus should have a PMU placed on it or be connected to a neighboring bus that has a PMU installed on it. A $f(Y)$ matrix is thus constructed which will indicate the relevant connections between each bus and the PMU. If two buses are connected, then the corresponding entry in the matrix would be one, otherwise it would be zero. The formulation of $f(Y)$ is given below:

$$f(Y) = MY \quad (7)$$

The number of PMUs required for complete observability (while considering zero injection buses) for the IEEE-118 bus system is 29. They were placed on buses 3, 8, 11, 12, 17, 20, 23, 28, 34, 37, 40, 45, 49, 52, 56, 62, 65, 72, 75, 77, 80, 85, 86, 91, 94, 102, 105, 110, and 114.

## V. RESULTS

A total of 4,800 cases were generated for the summer load profile out of which 1,780 cases were insecure and the remaining 3,020 cases were secure. The number of operating conditions considered was 96 (i.e. $N_{oc} = 96$), while the number of contingencies simulated was 50 (i.e. $N_C = 50$). This database is then fed to the different ML algorithms to try and classify the security of the system. From the first plot in Fig. 7 we observe that performance of RF (in terms of accuracy) in absence of errors is better than the other algorithms, with DT following it closely. The DT built for this case had a size of 6 for all the simulated cases. As mentioned in Section II, the training and the testing data set were split in the ratio of 70:30 and a 10-fold cross validation was performed for all the simulated cases. Different numbers of layers in the MLNN model with different activation functions were tested to find the optimal performance among them. The optimal number of layers for MLNN in this study was found to be 5.

In presence of errors, the performance of SVM was superior to the other techniques, with MLNN following it closely. Due to the nonlinear nature of the RBF kernel used in SVM, it was able to achieve higher levels of accuracy even when errors were introduced into the measurements. Note that although both PMU errors as well as instrumentation errors were considered in the study, the results were mostly dominated by the instrumentation errors, which have often been neglected in previous DSA studies.

4,800 cases were generated for the other seasons as well out of which 1,320 cases for spring, 1,510 cases for fall, and 1,392 cases for winter were insecure. The performances of the ML algorithms have also been validated across the other seasons as shown in Fig. 7. The performances of the algorithms have been found to be similar, i.e., RF has the highest accuracy in absence of measurement errors whereas SVM has outperformed the other three algorithms in presence of errors.

In the next set of simulations, approximately 10% and 20% of the total generation was replaced with solar PV and the accuracy of the scheme was tested on the modified system for a summer seasonal load profile. CAISO currently has approximately 10% total solar penetration in their system, while it is expected that this percentage will double within the next decade. The simulations were run 75 times and a 95% confidence interval was computed for the test data while adding different measurement errors during each run. The mean accuracy obtained over all the 75 runs are presented in Fig. 8 and Fig. 9. From Fig. 8 (9), we observe that, for 10 (20) % solar penetration, the performance of RF is better than other algorithms in absence of errors. With the addition of errors, SVM has the highest accuracy and its performance is better than the other three algorithms.

The results that have been shown (in Figs. 7-9) describe the classification accuracy for the different ML algorithms with and without the addition of errors. The ML algorithms have been tested with different seasonal load profiles as well as with two different percentages of solar PV in the grid. The importance of performing a seasonal based load modeling have been emphasized. Furthermore, the significance of solar integration while performing DSA has been discussed.

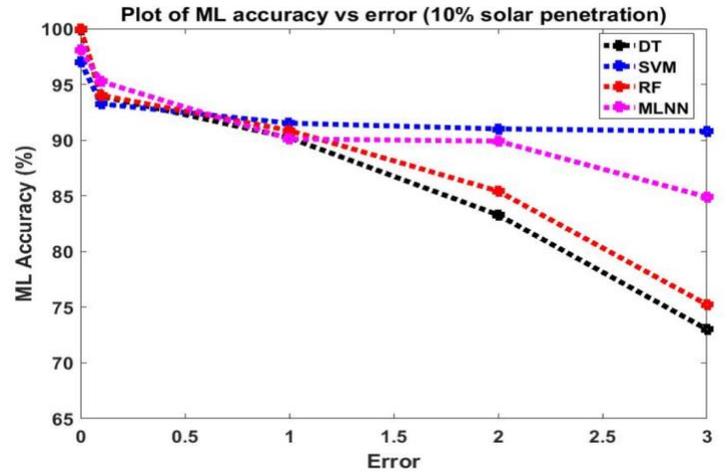
Fig. 8. Plot of ML accuracy vs error for 10% solar penetration

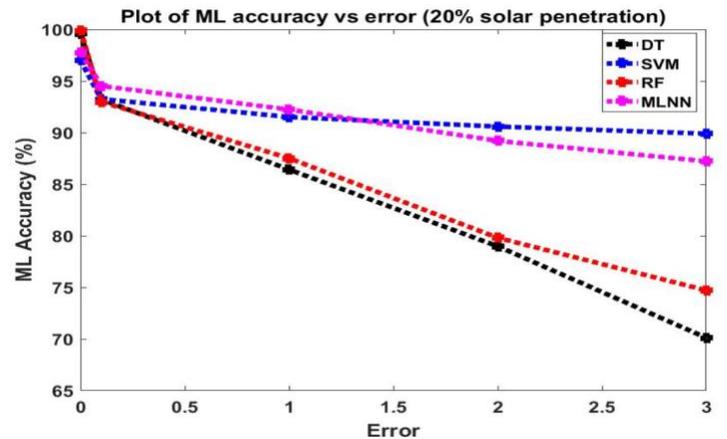
Fig. 9. Plot of ML accuracy vs error for 20% solar penetration

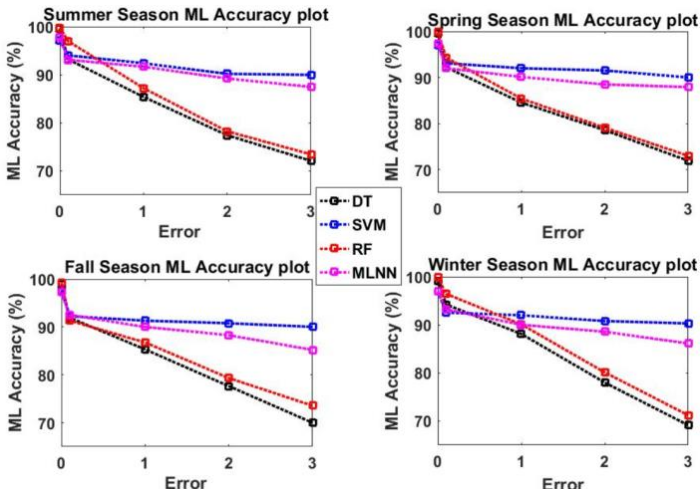
Fig. 7. Plot of ML accuracy vs error for different seasons

## VI. CONCLUSIONS AND FUTURE WORK

In this paper, a synchrophasor measurement based DSA scheme is developed considering different load variations corresponding to different seasons while also accounting for varying amounts of solar penetration and realistic errors in the PMU measurements. Different ML techniques have been employed, such as decision trees (DTs), support vector machines (SVMs), random forests (RFs), and multi-layer neural networks (MLNNs), to classify the security of the system under different conditions. Following conclusions can be drawn from the study:

a) The performance of RF was found to be the best among the three algorithms considered when measurement errors were not included in the study. Substantial degradation in performance of RF (and DTs) was observed when measurement errors (primarily the instrumentation errors) were introduced.

b) The performance of SVM and MLNN were affected to a lesser extent due to the presence of measurements errors.

c) The proposed scheme of using seasonal load has proved that under the same set of contingencies for a different season, the number of violations differ. Therefore, there is a need to include seasonal variability while doing DSA studies.

d) With the inclusion of renewables in the study, for the same contingency scenarios, the number of transient stability limit violations and voltage security limit violations increase.

During this study we came across the following scopes of research that can be explored in the future: a) effect of scalable loads can be incorporated into the study to make the load changes more dynamic in nature, b) use of ensemble learning techniques can be incorporated into the study for better classification accuracy, and c) performance on a hybrid renewable generation-rich system (having both solar as well as wind) can be investigated.